\documentstyle [11pt]{article}
\newcommand{\gsim}{\raisebox{-0.3ex}{\mbox{$\stackrel{>}{_\sim} \,$}}}
\newcommand{\lsim}{\raisebox{-0.3ex}{\mbox{$\stackrel{<}{_\sim} \,$}}}
\def\ni{\noindent}

\def\i{\item}
\def\cp{\clearpage}
\begin{document}
\small
\title{ A Study of the Orion Cometary Cloud L1616
\footnote{ Carried out under the  auspices of the Joint Astronomy  Program,
Department of Physics,  Indian Institute of  Science, Bangalore in  partial
fulfillment of the requirements for the Degree of Doctor of Philosophy.  }}
\author{B.Ramesh\\   Raman   Research   Institute,   \\   C.V.Raman Avenue,
Sadashivanagar, Bangalore 560080.\\ E-mail: bram@rri.ernet.in. \\}
\date{}
\maketitle
\begin{abstract}

    With  its  cometary  appearance  and  a  reflection nebula near its edge
facing some bright Orion stars, the Lynd's cloud L1616 shows ample  evidence
for being affected by one or  more of these massive stars.  To  estimate its
mass  and  star  formation  efficiency  as  well  as  to  determine if it is
gravitationally  bound,   we  mapped   this  cloud   in  J=1${\rightarrow}$0
transitions of $^{12}$CO and $^{13}$CO.   It is found that the  distribution
of the emission in the line {\it wings} show clear evidence for  substantial
mass motions.  Also, the  ``virial'' mass of the  cloud is found to  be five
times the  actual cloud  mass determined  from the  $^{13}$CO column density
map.   It  is  argued  that  this  cloud  has abnormally high star formation
efficiency and is possibly disintegrating.  The morphology and the  location
of the cloud  indicate that it  is being affected  by the star  ${\epsilon}$
Orionis which  is also  possibly responsible  for the  cloud's unusual  star
formation efficiency.  Over  a range of  values of the  relevant parameters,
the star is  found to quantitatively  satisfy the requirements  of being the
cause of the observed characteristics of the cloud.

\end{abstract}

\centerline{ Keywords: stars: formation; ISM: clouds; ISM: individual:
L1616; reflection nebulae. }
\vspace {.5 cm}
\centerline{ Running title: Cloud L1616 }
\vspace {.5 cm}
\centerline{ Proofs to: B.Ramesh}
\normalsize
\cp

\section {Introduction}

    Massive  stars  affect  the  structure  and  the evolution of the clouds
around  them  substantially.   Globules  and  clouds  with  bright  rims and
cometary appearance are found in the vicinity of many nearby OB associations
e.g.  Vela (Gum nebula; Sridharan 1992), Orion (Bally et al. 1991),  Cepheus
(Indrani \& Sridharan 1994) and Rosette (Patel, Xie \& Goldsmith 1993).   In
addition  to  affecting  the  morphology,  these  stars  also accelerate the
globules and possibly induce star formation in them as well.  There are many
clouds in the  Orion complex which  show evidence of  being affected by  the
nearby stars.  Lynd's dark cloud  L1616 with its cometary appearance  is one
among them.  It also harbors a bright reflection nebula NGC 1788, also known
as CED 040, excited by a poor star cluster.

    Study  of  a  cloud  with  an  associated  reflection nebula offers many
advantages.   Since  reflection  nebulae  mark  close spatial association of
relatively dense interstellar clouds with luminous stars of spectral type B1
or later, the distance to the star  and hence to the cloud can be  estimated
to a reasonable  accuracy.  This allows  one to determine  sizes, masses and
luminosities reliably.  A reflection  nebula with newly born  stars provides
an  oppurtunity  to  study  the  effects  of  recent formation of stars with
intermediate masses.  In particular, one that harbors a new-born cluster  is
most suited to study cloud fragmentation.  Stars, newly formed or otherwise,
have significant effects on the thermal balance of a cloud.  Hence, study of
such a cloud where there is an identifiable dominant energy source (but  not
as disruptive as O stars) permits one to investigate the heating and cooling
in them.

    L1616 appears to  be unique among  the reflection nebulae.   As shown in
the next section, L1616 has an estimated star formation efficiency of $\sim$
14\%, much larger than other clouds with associated reflection nebulae.  The
cometary appearance of the cloud,  the peculiar location of the  nebula near
its edge facing  the bright Orion  belt stars, and  its suspected high  star
formation efficiency  indicate that  the cluster  formation in  the cloud is
possibly induced.  Furthermore,  no IRAS source  with a spectrum  typical of
young stellar objects is found within the cloud boundary.  This may  suggest
that the formation of the cluster has pre-empted any further star  formation
- a self  regulatory process often  invoked in the  literature.  Thus, L1616
seems to be an object of considerable interest.  We mapped this cloud in J =
1 $\rightarrow$ 0 transitions of both $^{12}$CO and $^{13}$CO with a view to
determine its mass more reliably and to discern any mass motions that may be
present.  In  this paper  we present  the maps  and argue  that induced star
formation is most likely to be the case and that the cloud is fragmenting.

    In the  following section  we estimate  the star  formation efficiencies
(SFE) of a few clouds with associated reflection nebulae and show that L1616
is  unique  among  them.   In  Section  3  we  present  the  details  of the
observations and the maps of L1616.  The results are discussed in Section 4.

\section {SFE of clouds with reflection nebulae}

    Star formation efficiency (SFE) usually refers to the ratio of the total
mass in the stars to the sum of this stellar mass and the mass of the  cloud
or cloud  segment they  are born  in.  We  estimate the  masses in these two
components for  the bright  reflection nebulae  in the  following way  using
their optical and infrared data from the literature.

\begin {description}

    \item [{\bf Masses of the clouds : } ] { The cloud mass can be expressed
as  2$\times  N(H_{2})\,m(H)\,\Omega_{C}\,d^{2}$,  where  $N(H_{2})$  is the
molecular  hydrogen  column  density,  $m(H)$  the  atomic mass of hydrogen,
$\Omega_{C}$ the solid angle of the cloud and $d$ its distance from the Sun.
To estimate  $\Omega_{C}$ and  $N(H_{2})$ reliably,  the clouds  need to  be
mapped both in $^{12}$CO and $^{13}$CO.  However, for most of the reflection
nebulae  such  maps  are  not  available.   Hence,  we  use  their   optical
obscuration solid  angles listed  in the  catalogues of  dark clouds  (Lynds
1962; Feitzinger \& Stuwe 1984; Hartley et al. 1986).  We wish to note  that
only the mass in dense portions of the cloud, which the optical  obscuration
would  trace  well,  may  be  relevant  for the estimation of SFE. Molecular
hydrogen column density seems  to be nearly constant  in clouds over a  wide
range of sizes (Larson 1981) and we take this to be $\sim 5 \times  10^{21}$
cm$^{-2}$, typical of the small  molecular clouds.  The masses estimated  in
this way will be very approximate.  Nevertheless, we believe that this  will
not be a severe handicap for the relative comparison of SFE attempted here.}

    \item [{\bf Masses  in stars :  } ] {  Many optically bright  reflection
nebulae are  also bright  in far  infrared (FIR)  and have  been detected by
IRAS.  Their FIR  emission results from  reprocessing of the  radiation from
the exciting stars by the dust in the surrounding cloud.  In most cases, the
spectral type of the brightest member  of the exciting stars is also  known.
Then, if  one assumes  all stars  to be  of the  same spectral type, a lower
limit      to      the      mass      in      stars      is      given    by
$4\;\pi\;d^{2}\;S_{FIR}\;M\;\left(\eta\;L\right)^{-1}$.  Here  $M$ and  $L$,
respectively,  are  the  mass  and  luminosity  of the brightest star in the
cluster derived from its spectral type (Allen 1976), $\eta$ is the  fraction
of  the  stellar  radiation  that  is  reprocessed  which is taken to be 0.3
(albedo = 0.7, Witt  \& Schild 1986) and  $S_{FIR}$ is the FIR  flux density
derived in the following way using  the measured flux densities in the  four
IRAS bands (Margulis, Lada \& Young 1989):

\small
\begin{eqnarray}
S_{FIR} &=& \int\;S_{\nu}\;d\nu   \\   &=&   2.37   \times
10^{-10}\left(S_{12}   +    0.567\;S_{25}   +    0.154\;S_{60}   +
0.059\;S_{100}     +      \frac{7.39\;S_{\lambda_{max}}}{\lambda_{max}}\right)
ergs^{-1}cm^{-2} \nonumber
\end{eqnarray}
\normalsize

    In this expression the flux densities are in Jy and $\lambda_{max}$, the
longest wavelength at which the flux  density is measured is in $\mu$m,  and
$S_{\lambda_{max}}$ the corresponding  flux density.  The  last term is  the
estimated  flux  emitted  longward  of  $\lambda_{max}$ (Myers et al. 1987).
This is obtained  by assuming that  $S_{\lambda_{max}}$ is also  the maximum
over the entire spectrum, and that the spectrum is like that of a  blackbody
for wavelengths longer  than $\lambda_{max}$.  However,  it should be  noted
that  if  star  formation  is  {\it  very}  recent,  as  the luminosities of
protostars change considerably with time (Lada 1991), the above estimate  of
mass in stars can be inaccurate. } \end {description}

    Since  the  masses  of  both  the  components  depend  similarly  on the
distance,  their  ratio  is  independent  of  it  resulting in the following
expression for the minimum SFE.

\small
\begin{eqnarray}
SFE~(\%)~&=&~100~\left(\frac{X}{1+X}\right) \nonumber \\
\ni {\rm where,~~} \nonumber \\
X~&=&~\frac{M_{*}}{M_{C}} \nonumber \\
&=&~\left[\frac{4\pi\;S_{FIR}\;M}{\eta\;L}\right]          \times
\frac{1}{(2\;m_{H}\;N_{H_{2}}\;\Omega_{C})} \nonumber \\
{}~&=&~   1.286    \times   10^{9}~\frac{(M/M_{\odot})}{(L/L_{\odot})}    \;
\frac{(S_{FIR}/erg\;s^{-1}\;cm^{-2})}{(\Omega_{C}/\mu~\rm steradian)}
\end{eqnarray}
\normalsize

    \ni Thus,  the only  quantities entering  the estimates  are: the  solid
angle of the clouds, the IRAS  flux densities of the associated nebulae  and
the mass and the luminosity of the exciting star with the earliest  spectral
type.  We use this approach to determine SFE of a few clouds with associated
reflection nebulae.  In Table 1 we have listed the names of the nebulae (row
1), of the brightest stars exciting  them (row 2), their spectral type  (row
3), mass (row 4) and luminosity (row 5), the names of the associated  clouds
(row 6), their solid angle (row 7) and mass (row 8), names of the associated
extended IRAS sources (row 9), their  flux densities in the four bands  (row
10-13), the FIR fluxes $S_{FIR}$ (row 14), and their estimated SFE (row 15).
The spectral type of the stars exciting the nebulae have been taken from the
catalogues  of  reflection  nebulae  (van  den  Bergh 1966; van den Bergh \&
Herbst 1975).  Table 1 shows that  L1616 has the highest estimated SFE.  Two
other clouds, L1630 and SC064, also have high estimated SFE but their masses
are likely to be under-estimates owing to their large physical sizes.   This
can  be  seen  from  the  following.   One  could take the volume density of
molecular hydrogen, n$_{H_{2}}$, in the clouds to be the same instead of the
column density.  Then, for spherical clouds, the new SFE$_{V}$ is given by,

\small
\begin{eqnarray}
SFE_{V}~(\%) ~&=&~100~\left(\frac{X_{V}}{1+X_{V}}\right) \nonumber \\
\ni {\rm where,~~} \nonumber \\
X_{V}~&=&~1470  \times ~\frac{X}{(d/pc)~
(\sqrt {\Omega_{C}/ \mu~\rm steradian})}
\end  {eqnarray}
\normalsize

    \ni A value of 1100 cm$^{-3}$ has been assumed for n$_{H_{2}}$ such that
SFE and SFE$_{V}$  are the same  for the cloud  L1616 (This choice  does not
affect the relative values of SFE$_{V}$  derived).  Row 16 in Table 1  lists
SFE$_{V}$ calculated  this way  using distances  (last row)  taken from  the
literature (Racine 1968; Herbst 1975).  Table 1 shows that SFE$_{V}$ for the
two  clouds,  L1630  and  SC064,  are  much  smaller  than  that  of  L1616.
Nevertheless, they also appear  peculiar and merit further  studies.  Before
leaving this section, we wish to emphasise that the SFEs calculated here are
very approximate.   The estimated  stellar mass  being a  minimum makes  the
calculated SFE  to be  a lower  limit.  But,  this is  made quite  uncertain
because of the error  in the estimate of  the cloud mass.  For  example, the
assumption that the density is the same for all the clouds may not be valid.
In this sense, both the small  clouds (L1616 \& SG033) having high  SFEs may
be  a  bias  effect  resulting  from  them  being denser.  Thus, detailed CO
observations are  needed to  confirm these  SFEs.  We  have carried out such
observations for the cloud L1616.

\section {Observations}

    Above  estimates  suggest  that,   among  the  clouds  with   associated
reflection nebulae, the cloud L1616 has an abnormally high SFE. However,  as
mentioned already, it is necessary to make detailed CO maps to determine the
cloud mass and hence  the SFE more reliably.   Such maps would also  help to
ascertain if  mass motions  are present.   For this  purpose, we  mapped the
cloud  L1616  in  $J\,=\,1\,\rightarrow  0$  transitions  of  $^{12}$CO  and
$^{13}$CO with a $\sim\,1'$ beam and a grid-point spacing of one beam.   The
observations were carried  out during the  winters of 90-91  and 91-92 using
the 10.4m millimeterwave telescope  at the Raman Research  Institute campus,
Bangalore  (for  a  brief  description  of  the  telescope  see  Patel 1990;
Sridharan 1993).  A filter-bank spectrometer with 250 kHz (0.65  kms$^{-1}$)
resolution  covering  a  total  bandwidth  of  64  MHz was used.  During the
observations,  pointing  was  checked  by  beam  switched continuum scans on
Jupiter  (see  Patel  1990  for  details)  and  the  rms  pointing error was
estimated to  be $\sim\,12''$.   An ambient  temperature load  was used  for
calibration.  During the observations the DSB T$_{sys}$ ranged from 600$\,$K
to 1200$\,$K.   Frequency switching  by 15.25$\,$MHz  was used  for all  the
observations and  the spectra  obtained were  appropriately combined.  Third
order polynomials were fitted to  estimate and remove the curved  baselines.
Most of the final spectra had an rms noise of $\sim$0.25 K and the estimated
rms  error  on  the  velocities  is  $\sim$ 0.3 kms$^{-1}$.  We estimate the
forward spillover and  scattering efficiency, $\eta_{fss}$,  to be 0.57  and
0.63 at the $^{12}$CO and $^{13}$CO frequencies, respectively, by  comparing
the measured antenna temperatures on the calibration sources (mostly Ori  A)
with the source brightness temperatures reported in the literature (taken to
be 73 K and 16 K at  the two frequencies, respectively, for Ori A).  We have
also corrected the spectra for the elevation dependence of $\eta_{fss}$.

    The kinetic temperature and  the $^{13}$CO column density  were obtained
from the spectra using the relations given by Dickman (1978), assuming  that
$^{13}$CO is  optically thin,  and that  both $^{12}{\rm  CO}$ and $^{13}$CO
have the same  excitation temperatures.  These  relations take into  account
the  2.7  K  microwave  background  as  well  as  the  fact  that  the usual
Rayleigh-Jeans approximation  ($h \nu  < <  kT$) is  not valid at millimeter
wavelengths  for  the  typical  kinetic  temperatures  that  obtain  in  the
molecular clouds.  Figures  1 to 4  show the distributions  of the $^{13}$CO
column  density,  the  kinetic  temperature,  the  equivalent  width and the
integrated line intensity ($\int\;T\;dV$) in the cloud L1616.  The $^{12}$CO
spectra were used  to obtain the  last three quantities  while the $^{13}$CO
spectra yielded the first.   The ellipse in Figs.$\,$1  to 4 outlines the  I
band image  of the  nebula NGC  1788 (see  Plate 62  of Witt \& Schild 1986)
associated  with  L1616.   Its  major  and  minor  axes are $\sim\,3.4'$ and
$\sim\,2.1'$ long, respectively.  It is centred at the position of the  IRAS
source  with  $RA  =  5^{h}\;4^{m}\;25.8^{s}$  and  $DEC = - 03^{\circ}\;25'
\;5''$ ( hereafter, the {\it centre}) and oriented $45^{\circ}$ with respect
to the west.   The star, HD  293815, of spectral  type B9V is  the brightest
visible member of the cluster of stars illuminating the nebula NGC 1788.  It
lies $\sim\,0.8'$ away from the {\it centre} in the NW direction.

Following are the notable features in the maps:

\begin{enumerate}

    \item {The $^{13}$CO column density  peaks at $\sim3'$ west of  the {\it
centre} with a {\it shoulder} around the nebula and falls off gradually  and
more or less symmetrically from the peak.}

    \item {The  temperature distribution  is more  or less  centered on  the
nebula  with  its  peak  at  $1'$  east  of  the  {\it centre} and hence not
coincident with the column density peak.  This indicates that the  radiation
from the stars is the dominant energy source.}

\item  {The  spatial  distribution  of  equivalent  widths  is more or less
symmetrically  distributed  and  has  its  peak  coincident with the column
density peak.   Fig.3 shows  that most  parts of  the cloud have equivalent
line widths greater than $\sim2\;{\rm km\;s}^{-1}$.}

    \item {The  contour map  of the  derived {\it  integrated $^{12}$CO line
intensity} distribution in L1616  bears similarity to both  $^{13}$CO column
density (Fig.1) and kinetic temperature (Fig.2) distributions.  For example,
the two peaks, one each to the east and west of the {\it centre}, correspond
to the temperature and column density peaks, respectively.  This  similarity
implies that although the $^{12}$CO integrated line intensity map has column
density information, it is substantially altered by the distribution of  the
temperature  in   the  cloud.    Hence,  using   integrated  $^{12}$CO  line
intensities to trace the column density distributions of molecular  hydrogen
is not always reliable.}

\end{enumerate}

\section {Discussion}

    As stated earlier, one of the primary objectives of the observations was
to determine the cloud mass more reliably.  The mass of the cloud  estimated
from the  $^{13}$CO column  density map  presented in  the previous  section
using a N$_{\rm H_{2}}$ to N$_{13_{\rm CO}}$ conversion factor of $5  \times
10^{5}$  (Dickman  1978)  is  $\sim169\;M_{\odot}$.   One  can  also use the
integrated $^{12}$CO line  intensity map to  determine the cloud  mass.  The
mass found this way using a value of $2.3 \times 10^{20}\;{\rm cm}^{-2}/{\rm
K\;km\;s}^{-1}$ for the I$_{12_{\rm  CO}}$/N$_{\rm H_{2}}$ ratio (Strong  et
al.  1988)  is  $\sim193\;M_{\odot}$.   The  two  mass estimates are in good
agreement   with   each   other.     We   take   their   mean,    which   is
$\sim180\;M_{\odot}$, to be the cloud mass.  This agrees well with our rough
estimate  presented  in  the  beginning,  although  larger  by  a  factor of
$\sim\,$1.15.   The  SFE  calculated  using  this  new  and  more   reliably
determined  cloud  mass   is  $\sim$12\%  (estimated   total  mass  of   the
stars/stellar and cloud mass).  This is still large compared to the  average
SFE value of $\lsim$ 3\% (Evans \& Lada 1991) found for the nearby molecular
clouds.

    One  can  also  obtain  the  mass  from  the  cloud-averaged line width,
provided the cloud is in virial  equilibrium.  For L1616, since the line  is
wide over  most parts  of the  cloud, the  actual cloud-averaged  line width
turns  out  to  be  $\sim\,2.7\,$kms$^{-1}$.   The corresponding virial mass
required to  keep the  cloud bound  is $1000\;M_{\odot}$,  five times larger
than the estimated  mass.  Conversely, if  the cloud is  bound, its detected
mass of $\sim180\;M_{\odot}$ would imply a cloud-averaged line width of only
$\sim1.25\;{\rm km\;s}^{-1}$.  Thus, it  appears that the energy  input from
the  stars  in  the  cluster  is  {\it  fragmenting}  the cloud.  The excess
turbulent motions of $\sim\,1.5\,$kms$^{-1}$ over and above that needed  for
virial equilibrium and the present cloud size of $\sim\,2\,$pc suggest  that
the fragmentation is possibly happening for the past 1 to 2 Myr.

    The spatial distributions  of emission in  the {\it linewings}  shown in
Fig.5 also indicate the disturbed state of the cloud.  The average  emission
from the material in  the velocity ranges 5.4  to 6.7 kms$^{-1}$ (thin)  and
8.7 to  10.4 kms$^{-1}$  (thick), respectively,  are shown  superimposed one
over the other.  The {\it  blueshifted} and {\it redshifted} emissions  have
large spatial extents and distinct peaks,  each a little away from the  {\it
centre}.  They  clearly indicate  that substantial  mass motion  is present,
possibly caused by the activity of  the stars in the cluster.  For  example,
the spherical {\it redshifted} emission,  whose peak is coincident with  the
column density  peak, is  found to  contain $\sim\,$10\%  of the cloud mass.
The conclusion that the mass motions in the cloud is substantial is  further
supported by the coincidence of the column density peak with the line  width
peak.  It is also consistent with the conclusion of de Vries et al.   (1984)
who, from a systematic survey of reflection nebulae, found that 60\% of  the
clouds with  associated reflection  nebulae have  broad lines;  the enhanced
emission in the wings and local line broadening arise from mass motions  due
to  the  dynamical  interaction  between  the  molecular cloud and the stars
illuminating the nebulae.   Incidentally, the energy  contained in the  mass
motions is  $\sim\,10^{45}$ ergs  which is  less than  $\sim\,$0.1\% of  the
energy radiated by the stars in the cluster over the past few million years.

Having found that the cloud L1616  has abnormally high SFE and shows  clear
signs of disintegration, we set  out to determine the possible  cause.  Its
high  SFE  and  cometary  appearance  suggest  an external cause.  Figure 6
presents a wide angle view of the portion of the sky around the cloud L1616
showing its  cometary tail  towards $30^{\circ}$  southwest and  all nearby
stars of  spectral type  earlier than  B2. Only  such stars  can affect the
morphology and kinematics of nearby  clouds significantly as well as  cause
implosion in  them.  Table$~$2  provides the  details of  these stars taken
from the  literature (Hirshfeld  \& Sinnott  1985).  The  projected tail is
directed  nearly  perpendicular  to  the  galactic  plane and its extension
points to stars 3 and 4 as possible causes.  However, when their  distances
from the cloud and  their spectral types are  considered, it is clear  that
star 3 ($\epsilon$ Orionis, a blue supergiant) is the most influential one.
Its present surface temperature and luminosity estimated from its  spectral
type  (B0Ia)  are  $\sim$  28000$\,$K  and  $2.5 \times 10^{5}\;L_{\odot}$,
respectively (Allen  1976).  This  implies that  the spectral  type of  its
main-sequence (MS) progenitor is O6  with a mass of $\sim$  $35\;M_{\odot}$
and a  lifetime of  $\sim$ 4.3  Myr. Such  a star  may have accelerated the
cloud through the {\it  rocket effect} (Oort \&  Spitzer 1955), as well  as
causing  it  to  implode  and  form  the star cluster.  Earlier works (e.g.
Bertoldi 1989, Bertoldi \& McKee 1990) have shown that a cloud can  implode
as well as acquire a tail  structure in such a process.  A  simple estimate
of the time taken by the  cloud to cover the current star-cloud  separation
of  50-70  pc  using  the  typical  induced  velocities of 10-15 kms$^{-1}$
(Bertoldi and McKee 1990) yields a value of $\sim\;5\; $Myr.  This being in
good  agreement  with  the  expected  age  of  the  star $\epsilon$ Orionis
suggests that its  progenitor could have  propelled the cloud  L1616 to its
present position  from an  initial location  close to  the star  within the
stellar life-time.  In the following,  we proceed to check this  suggestion
more quantitatively.

Using  two  dimensionless  parameters  viz.  the  column density parameter,
$log(\eta)$, and the initial shock velocity parameter, $log(\nu)$, Bertoldi
(1989)  has  classified  clouds  exposed  to ionising radiation.  The cloud
L1616, given its radius of $\sim\,1\,$pc and molecular hydrogen density  of
1200$\,cm^{-3}$, has a  value of $\sim\,3.24$  for $log(\eta)$.  The  upper
and lower bounds to  its initial distance of  50$\,$pc (set by the  present
projected cloud separation) and 1$\,$pc, respectively, constrain $log(\nu)$
to lie between -0.62 to 1.08.  These values for the parameters place  L1616
in  region  II  of  his  Fig.1.   In  this case, almost all of the ionising
radiation  will  be  absorbed  by  the  recombining  gas  streaming off the
ionisation front and the ionisation  front can be considered thin  relative
to the cloud  size.  In this  approximation, the temporal  evolution of the
parameters  (mass,  radius,  ion  density,  distance  and speed) of a cloud
exposed to  radiation from  a massive  star are  described by the equations
presented  in  Appendix  A.  There  are  three unknowns in the problem: the
escape  velocity  of  the  ionised  plasma,  $v_{i}$,  the fractional cloud
velocity at present, $\frac{v_{1}}{v_{i}}$ and the ratio of the initial and
the final  values of  the cloud  distance, $\frac{x_{0}}{x_{1}}$.  However,
the latter two are not independent  (as can be seen from equation  A4) thus
reducing the independent unknowns to two.  The only constraint is that  the
cloud moves from $x_{0}$ to  $x_{1}$ within the interaction time-scale  set
by the age of the star.  This  is bound by its MS life-time of  $\sim\,$4.3
Myr and its maximum age of $\sim\,$5 Myr.

Fig.7  shows  a  contour  plot  of  the  time needed to move to the present
separation, taken  to be  50$\,$pc, as  a function  of $x_{0}$ and $v_{i}$.
For the same value of $v_{i}$ two initial positions, one closer to the star
and another closer to  the cloud, result in  the same time of  travel.  The
former solution  is preferred  as it  results in  a larger  average density
enhancement    ahead    of    the    shock    front    (which    goes    as
$ln(\frac{x_{1}}{x_{0}})$),  thus  increasing   the  chances  of   external
triggering.  Although, $x_{0}$ and $v_{i}$ can not be uniquely fixed, it is
clear that, for  reasonable values of  these parameters, the  progenitor of
the star $\epsilon$ Orionis could  have caused the propulsion of  the cloud
to its present position within the stellar life-time.  For a temperature of
10000$\,$K, the sound speed in the ionised plasma is $\sim\,11.4\,kms^{-1}$
and, observationally, the plasma in  a bright condensation in M16  is found
to be streaming with  a speed of $\sim\,13\,kms^{-1}$  (Courtes, Cruvellier
\& Pottasch, 1962).  Hence, we take $v_{i}$ to be typically between 12  and
15 kms$^{-1}$.  The projected and  3-D separation between the star  and the
cloud are 50$\,$pc  and 70$\,$pc, respectively.   If these are  taken to be
the  lower  and  upper  limits  to  the  distance  travelled in the maximum
available time of $\sim\,5\,$Myr,  along with a corresponding  escape speed
for the plasma of 12 and 15 kms$^{-1}$, the needed initial separations turn
out to be 5$\,\,$pc and 4$\,\,$pc, respectively.  Fig.8 shows the variation
of the cloud parameters with  time for these two cases.   The corresponding
initial mass, initial radius and the present speed of the cloud for the two
cases   are:   $535\;$M$_{\odot}$,   1.29$\,$pc,   13.1$\,$kms$^{-1}$   and
$595\;$M$_{\odot}$, 1.34$\,$pc, 18.0$\,$kms$^{-1}$, respectively.  For  the
lower    velocity    case,     the    measured    radial     velocity    of
$\sim\,7.5\,$kms$^{-1}$, which  includes a  galactic differential  rotation
contribution  of  $\sim\,3.7\,$kms$^{-1}$,  would  imply  a  proper  motion
velocity  of  $\sim\,12.5\,$kms$^{-1}$.    This  would  require   that  its
direction of motion makes an angle $\gsim\,73^{\circ}$ with respect to  the
line of sight (LOS) or, equivalently, the difference in the distance to the
star and the cloud from the Sun is not more than $\sim\,$15$\,$pc.  This is
within the errors  ($\sim$ 20\%) of  the estimated distances  to $\epsilon$
Orionis and L1616  of 370 pc  and 420 pc,  respectively.  Incidentally, the
density $\rho_{2}$ behind the shock  front varies from 150 to  15 cm$^{-3}$
(see Fig.8).  The density ahead of the front $\rho_{1}$ is approximately,

\small
\begin{eqnarray}
\rho_{1} &=& 2\;\rho_{2}\ \left(\frac{v_{2}}{v_{1}}\right)^{2} \nonumber \\
&=& 4\;\rho_{2}\ \left(\frac{T_{2}}{T_{1}}\right)
\end  {eqnarray}
\normalsize

\ni  where,  $v$'s  and  $T$'s  are  the  corresponding  sound  speeds  and
temperatures.   Taking  $T_{1}$  and  $T_{2}$  to  be  40  K  and  10000 K,
respectively, $\rho_{1}$ is  found to vary  from $10^{5}$ to  $10^{4}$ over
the last 5$\,$Myr.  This  substantial density enhancement in  the frontside
of the cloud over  a long period could  have caused it to  implode and form
the  cluster.   The  cloud  fragmentation  timescale  estimated  earlier is
consistent with this scenario.   A quantity which could  have substantiated
this further is  the tail-stretching timescale.   However, it could  not be
obtained as no velocity gradient along the tail was discernible.  This  may
be  due   to  the   insufficient  velocity   resolution  used  ($0.65\;{\rm
km\;s}^{-1}$  at  115  GHz).   Also,  it  may be difficult to determine any
possible gradient due to the  confusion resulting from the mass  motions in
the cloud generated by the  cluster.  Nevertheless, assuming that the  tail
makes an angle of $\sim\,73^{\circ}$ with the LOS and the projected  length
of the tail  is $\sim\,2.5\,$pc, the  upper limit to  the velocity gradient
can be translated  to a lower  limit to the  tail stretching time-scale  of
$\sim\,1.1\,$Myr.  This agrees with the available travel time, but does not
constrain it well.  In conclusion, the above discussion is consistent  with
the  hypothesis  that  the  star  $\epsilon$  Orionis has been the external
trigger.  A  measurement of  the proper  motion velocity  of HD293815,  the
brightest star exciting the reflection nebula, will confirm this.

\section {Summary}

    We shall now summarise our  main findings.  Our main objectives  were to
determine the mass of  the cloud L1616 and  to find out if  large scale mass
motions are present.  For  this purpose, we mapped  this cloud in the  J = 1
$\rightarrow$  0  transitions  of  both  $^{12}$CO  and  $^{13}$CO.  We have
presented the  distributions of  the $^{13}$CO  column density,  the kinetic
temperature, the equivalent width and  the integrated line intensity in  the
cloud.  They clearly indicate that the star cluster exciting the  reflection
nebula is the dominant energy  source and that substantial mass  motions are
present in  the cloud.   The masses  determined from  the maps  of $^{13}$CO
column density and $^{12}$CO integrated line intensity are in agreement with
each   other   and   suggest   that   the   mass   of   the   cloud L1616 is
$\sim180\;M_{\odot}$.   This  confirms  the  high  star formation efficiency
suspected in this cloud if the stellar mass estimated from the FIR fluxes is
reasonably  correct.    However,  the   virial  mass   estimated  from   the
cloud-averaged line width of $\sim3\;{\rm km\;s}^{-1}$ is $1000\;M_{\odot}$.
This is much larger than the above estimates and suggests that the cloud may
be  disintegrating.   The  evidence  for  mass  motions  supports this.  The
cometary appearance, the location of  the star cluster near the  edge facing
the bright  Orion stars,  and the  estimated high  star formation efficiency
suggest that the cluster formation may have been externally triggered.   The
star {\it $\epsilon$ Orionis} located opposite to the tail direction is  the
most  likely  external  trigger.   It  is  found  that  this star could have
propelled the cloud from its initial  distance to its present position in  a
time-scale of $\sim\,$5 Myr for reasonable values of the two parameters viz.
the initial distance to the cloud  from the star and the escape  velocity of
the  ionised  gas.   If  the  cluster-formation is externally triggered, the
average density  enhancement must  have been  high.  This  requires that the
cloud was initially close to the star which leads to a large space  velocity
of $\sim\,13\,$kms$^{-1}$  at present.   Then, the  measured radial velocity
requires  the  cloud  to  have  a  proper  motion velocity of $\sim 12\;{\rm
km\;s}^{-1}$.

%\cp
\vspace{1.0cm}
\ni {\large \bf Acknowledgements}

    I am grateful to the faculty of the Joint Astronomy Program,  Department
of  Physics,  Indian  Institute  of  Science,  Bangalore for providing me an
opportunity to do research in astronomy.   My sincere thanks are due to  the
staff of  the Millimeterwave  Laboratory and  the Observatory  at the  Raman
Research Institute for their help and support during the observations.  I am
also  thankful  to  K.R.Anantharamiah  and  T.K.Sridharan for their critical
comments  on  the  manuscript  and  to  the anonymous referee for the useful
suggestions and critical comments.

\vspace{1cm}
\cp
\ni {\large \bf References}
\small
\begin{description}
\i { Allen C.W., 1976, Astrophysical Quantities.  William Clowes and  Sons,
London, p. 209 }
\i { Bally J., Langer W.D.,  Wilson R.W., Stark A.A., Pound M.W.,  1991, in
Falgarone  E.,  Boulanger  F.,  Duvert  G.,  eds,  Proc.   IAU  Symp.  147,
Fragmentation of Molecular Clouds  and Star Formation.  Kluwer,  Dordrecht,
p.11}
\i { Bertoldi F., 1989, ApJ, 346, 735 }
\i { Bertoldi F., McKee C.F., 1990, ApJ, 354, 529 }
\i { Courtes G., Cruvellier P., Pottasch S.R., 1962, Ann. d'Astroph., 25,
214}
\i  {  de  Vries  C.P.,  Brand  J.,  Israel F.P., de Graauw Th., Wouterloot
J.G.A., van de Stadt H., Habing H.J., 1984, A\&AS, 56, 333}
\i { Dickman R.L., 1978, ApJS, 37, 407 }
\i {  Evans$~~$II N.J.,  Lada E.A.,  1991, in  Falgarone E.,  Boulanger F.,
Duvert G., eds, Proc.  IAU Symp. 147, Fragmentation of Molecular Clouds and
Star Formation.  Kluwer, Dordrecht, p.293}
\i { Feitzinger J.V., Stuwe J.A., 1984, A\&AS, 58, 365 }
\i {  Hartley M.,  Manchester R.R.,  Smith R.M.,  Trittou S.B.,  Goss W.M.,
1986, A\&AS, 63, 27 }
\i { Herbst W., 1975, AJ, 80, 212 }
\i  {  Hirshfeld  A.,  Sinnott  R.W.,  1985,  Sky  Catalogue  2000.0 Vol.2.
Cambridge Univ.  Press, Cambridge }
\i { Indrani C., Sridharan T.K., 1994, JA\&A, 15, 157 }
\i { Lada C.J., 1991, in Lada  C.J., Kylafis N.D., eds, NATO ASI series  C,
342, The Physics  of Star Formation  and Early Stellar  Evolution.  Kluwer,
Dordrecht, p.329 }
\i { Larson R.B., 1981, MNRAS, 194, 809 }
\i { Lynds B.T., 1962, ApJS, 7, 1 }
\i { Margulis M., Lada C.J., Young, E.T., 1989, ApJ, 345, 906 }
\i {  Myers P.C.,  Fuller G.A.,  Mathieu R.D.,  Beichman C.A., Benson P.J.,
Schild R.E., Emerson J.P., 1987, ApJ, 319, 340}
\i { Oort J.H., Spitzer L., 1955, ApJ, 121, 6 }
\i {Patel N.A., 1990, Ph.D.thesis.  Indian Institute of Science, Bangalore}
\i { Patel N.A., Xie T., Goldsmith P.F., 1993, ApJ, 413, 593 }
\i { Racine R., 1968, AJ, 73, 223 }
\i { Spitzer L., 1978, Physical Processes in the Interstellar Medium. John
Wiley and Sons, New York, p. 262 }
\i { Sridharan T.K., 1992, JA\&A, 13, 217 }
\i { Sridharan T.K., 1993, Bull.Astron.Soc.India, 21, 339 }
\i { Strong A.W., Blomen J.B.G.M., Dame T.M., Grenier I.A., Hermsen W.,
Lebrun F., Nyman L.A., Pollock A.M.T., Thaddeus P., 1988, A\&A, 207, 1 }
\i { van den Bergh S., 1966, AJ, 71, 990 }
\i { van den Bergh S., Herbst W., 1975, AJ, 80, 208 }
\i { Witt A.N., Schild R.E., 1986, ApJS, 62, 839 }
\end{description}
\normalsize

\cp
\setcounter {section} {1}
\setcounter {equation} {0}
\renewcommand {\thesection}{\Alph{section}}
\renewcommand {\theequation}{\thesection \arabic{equation}}
\centerline {\large \bf APPENDIX: EVOLUTION OF CLOUD PARAMETERS}
\centerline {\large \bf UNDER ROCKET ACCELERATION}
\vspace {.5cm}

    \ni  An  early  type  star,  when  first  formed, is expected to pump UV
photons  into  the  surrounding  ISM  rather abruptly.  Many dense molecular
condensations within its Stromgren radius  (which is 12$\,$pc for a  star of
spectral  type  O6V  for  an  assumed  value of 10$\,$cm$^{-3}$ for the mean
hydrogen density of  the surrounding medium)  will experience a  weak R-type
ionisation front  passing quickly  around them,  after which  the stellar UV
radiation will drive an ionisation front into the denser cloud.  The ionised
hydrogen produced  on the  side of  the cloud  facing the  star, being  at a
higher pressure  than the  gas outside  owing to  its higher  density, would
expand producing a recoil on the cloud and thus accelerate it away from  the
star.  The loss of this gas also results in the reduction of the cloud mass,
and hence the cloud radius.  In this appendix we derive equations describing
how the  cloud's position  $x$, radius  $r$, mass  $m$ and relative velocity
with respect  to the  star $v$  change with  time under certain assumptions.
Let $v_{1},~m_{1}$ and $r_{1}$, respectively, be the present velocity,  mass
and  radius  of  the  cloud  and  $x_{0}$  and $x_{1}$, respectively, be its
initial and present distance from the star $\epsilon$ Orionis.   Integrating
the  equation  of  motion,  taking  $v$  to  be zero initially, one gets the
following result for rocket acceleration,

\small
\begin{eqnarray}
M(v) &=& M_{1}\ exp\left(\frac{v_{1}-v}{v_{i}}\right) \nonumber \\
 &=& M_{1}\ exp\left(z_{1}-z\right)
\end  {eqnarray}
\normalsize

    \ni Here, $z$ is $v$ expressed as a fraction of $v_{i}$, the velocity at
which the  ionised gas  escapes.  The  latter is  nearly equal  to the sound
speed in the ionised plasma which is $\sim\,$11.4 kms$^{-1}$.  In  obtaining
the above  result we  have ignored  the deceleration  due to  sweeping up of
matter.  A cloud of radius $\sim\,$1  pc ploughing through a medium with  an
average hydrogen  density of  1 cm$^{-3}$  over a  distance of $\sim\,$70 pc
would sweep up  only a mass  of $\sim\,$4 M$_{\odot}$.   Since this is  much
smaller compared to the  present cloud mass, deceleration  due to it can  be
ignored.  This is  particularly justified in  the case of  L1616 because the
cloud   is   at   an   altitude   of   $\sim\,-$190   pc   while   the  star
$\epsilon\,$Orionis is  at an  altitude of  $\sim\,-$120 pc  (Their galactic
longitudes  and  latitudes  are:  [203.5,  $-$24.7]  and  [205.2,  $-$17.2],
respectively).  Thus,  most of  its motion  would have  taken place  at high
altitudes, nearly perpendicular to the galactic plane, where the general ISM
density is likely to be quite low resulting in negligible swept-up mass.

    The   rate   at   which   the    cloud   loses   mass   is   given    by
$m_{p}\,n(z)\,\pi\,r^{2}(z)\,v_{i}$, where $m_{p}$ is  the mass of a  proton
and $n(z)$ is the number density of hydrogen ions just behind the ionisation
front (Spitzer 1978).  Here,  both $n$ and $r$  are taken to depend  on $z$.
As the cloud recedes from the star, the ionising photon density at the cloud
surface will vary and hence cause $n$ to vary.  Since the cloud loses  mass,
the cloud radius would also change.  If the mean density is conserved,  then
$r(z)$ is given by,

\small
\begin{eqnarray}
r(z)  &=& r_{1}\ exp\left(\frac{z_{1}-z}{3}\right)
\end  {eqnarray}
\normalsize

    \ni   Assuming   that   the   gas   expands   spherically   and that the
recombinations occuring in it nearly completely {\it shield} the stellar  UV
photons, one obtains the following expression for $n(z)$ (Spitzer 1978),

\small
\begin{eqnarray}
n(z) &=& \frac{1}{x(z)}~ \sqrt{\frac{3~N_{c}}{4\pi\alpha~r(x)}}
\end  {eqnarray}
\normalsize

    \ni where, $N_{c}$  is the stellar  output rate of  ionising photons and
$\alpha$ is the effective recombination probability for hydrogen taken to be
$2.6\,10^{-13}$  cm$^{3}$s$^{-1}$.   Now,  equating  the  mass  loss rate to
$\frac{M(z)}{v_{i}} \times \frac{dv}{dt}$ and  solving for $x(z)$ using  the
fact that $\frac{dz}{dt} \,=\, v_{i}\,z\, \frac{dz}{dx}$ one obtains,

\small
\begin{eqnarray}
x(z) &=&
x_{0}exp\left(\beta~\left[exp(0.5z_{1})-\left(1+0.5z\right)exp\left(
\frac{z_{1}-z}{2}
\right) \right] \right) \\
\ni {\rm where},~~
\beta~&=&~\frac{4M_{1}}{m_{p}}~\sqrt{\frac{4\alpha}{3\pi~N_{c} ~r^{2}_{1}}}
\end  {eqnarray}
\normalsize

    It is worth noting that, in addition to their boundary values, the cloud
variables described by the above equations depend only on $z$ and not on $v$
or $v_{i}$ themselves.  The fact that the infinitesimal time $dt$ taken  for
the cloud to move by an infinitesimal distance $dx$ depends on the  velocity
and acceleration during that period results in the following equation:

\small
\begin{eqnarray}
dt(z) &=& \frac{\beta~x(z)exp((z_{1}-z)/2)}{8v_{i}}~ \left[-z+
\sqrt{z^{2}+\frac{16~exp((z-z_{1})/2)~dx}{\beta~x(z)}}~ \right]
\end  {eqnarray}
\normalsize

    \ni Since it was not possible to obtain a closed form solution, we  have
numerically integrated this equation to get the time as a function of z.  In
the graphs shown in Fig.8, the time axis was derived in this way.

\cp
\small
\ni {\bf  Table  1.} Reflection nebulae $-$ star formation efficiencies \\
\begin{tabular*}{18.0cm} [t] {@{\extracolsep{\fill}}
lcccccccccc} \hline
 Nebula                         &  vdB017 &   vdB013 &   vdB033 &   vdB052
&   vdB055 &   vdB123 &   VHE05   &    VHE30   &  VHE17  &   VHE28  \\ \hline
 Star                           & +30$^{\circ}$549 &  +30$^{\circ}$540
&   293815 &    37903 &    38023 &   170634 &           &            &   &  \\
 SpT                            &  B8V    &   B8V    &   B9V    &   B1.5V
&   B4V    &   B7V    &  *        &  *         &  B1V    &   B5Vp           \\
 M$_{*}$        (M$_{\odot}$)   &  4.1    &   4.1    &   3.6    &   12.7
&    7.7   &    4.7   &    3.2    &     3.2    &  14.2   &    6.5           \\
 L$_{*}$        (L$_{\odot}$)   &  179    &   179    &   117    &   6887
&   1406   &    283   &     79    &      79    &  9727   &    794           \\
 Cloud                          &  L1450  &   L1452  &   L1616  &   L1630
&   L1641  &   L572   &    SC025  &      SC102 &   SG033 &    SC064         \\
 $\Omega_{C}$   (${\mu}$str.)   &  358    &   358    &    12    &   611
&   611    &  1322    &   171     &    764     &    8    &   212            \\
 M$_{C}$        (M$_{\odot}$)   &  6739   &   2227   &    157   &  13862
&   6617   &  18935   &  22279    &   52259    &   547   &   3991           \\
 IRAS                           &  X0326  &   X0322  &  X0504   &  X0539
&  X0539   &  X1827   &  X0809    &   X0916    & X0833   &  X0857           \\
 ~~~~name                       &  +312   &   +307   &   $-$034   &   $-$019
&   $-$081   &   +012   &   $-$356    &    $-$482    &  $-$405   &   $-$435 \\
 12${\mu}$m     (Jy)            &   91    &     6    &    30    &  2090
&    52    &    32    &     3     &     14     &  117    &   984            \\
 25${\mu}$m     (Jy)            &   30    &     6    &    68    &177700
&    53    &    65    &     4     &     28     &  190    &  4410            \\
 60${\mu}$m     (Jy)            &  1080   &     20   &    323   &  93400
&    304   &    375   &   14.5    &     196    &  1100   &  26400           \\
 100${\mu}$m    (Jy)            &  1190   &     49   &    594   &  77400
&    787   &    750   &     29    &     392    &  2700   &  24300           \\
 S$_{FIR}$      (10$^{-10}$)    &  1024   &     45   &    467   & 302246
&    553   &    536   &     27    &     266    &  1784   &  25530           \\
 SFE            (\%)            &  0.84   &   0.04   &  13.42   &  10.55
&   0.06   &   0.09   &   0.08    &    0.18    &  4.04   &  11.32           \\
 SFE$_{V}$      (\%)            &  0.13   &   0.01   &  13.42   &   1.23
&   0.01   &   0.01   &   0.01    &    0.01    &  2.20   &   2.47           \\
 Dist.          (pc)            &  501    &   288    &   417    &   550
&   380    &   437    &  1318     &    955     &  955    &   501    \\ \hline
\end{tabular*}
\vspace {0.2cm} \\
\ni {~~~~~~~A  (*)  mark  in  SpT  row  indicates  that the earliest stellar
spectral type is  not known and  has been assumed  to be A0V.  In cloud name
prefix L stands for  Lynd's cloud, SC for  southern dark cloud and  SG for
southern dark globule.} \vspace {1.0cm} \\

\ni {\bf  Table  2.} Luminous stars near the cloud L1616 \\
\begin{tabular*}{12.5cm} [t] {@{\extracolsep{\fill}}
rcclcc } \hline
 Index &SAO  &Dist. &SpT   &$\alpha_{1950}$  &$\delta_{1950}$     \\
 no.   &no.    &pc.   &    &h$~~$m$~~$s &$~~^{\circ}~~~'~~~''$ \\ \hline
  1    &132406 &500 &O9.5V &5 36 14 &$-$2 37  39   \\
  2    &132387 &540 &B1.5V &5 35 25 &$-$4 50  32   \\
  3    &132346 &370 &B0Ia  &5 33 41 &$-$1 13  56   \\
  4    &132269 &560 &B2V   &5 31 31 &$-$1 04  07   \\
  5    &132210 &550 &B2V   &5 28 55 &$-$6 44  41   \\
  6    &132222 &560 &B0V   &5 29 31 &$-$7 20  13   \\
  7    &112830 &560 &B1.5V &5 27 19 &+1   45  05   \\
  8    &112861 &470 &B1.5V &5 28 37 &+3   15  21   \\
  9    &112734 &490 &B1V   &5 22 09 &+1   48  08   \\
 10    &112697 &470 &B1V   &5 20 12 &+3   29  52   \\
 11    &131451 &460 &B2V   &4 41 41 &$-$8 35  44   \\ \hline
       &L1616  &420 &Cloud &5 04 30 &$-$3 25  00   \\ \hline
\end{tabular*}
\normalsize

\cp
\ni {\large \bf Figure Captions}

    \ni Fig.1~:~~Distribution  of $^{13}$CO  column density  in L1616.   The
ellipse marks the  position, the orientation  and the extent  of the I  band
image  of  NGC  1788.   The  peak  contour level is 3.2$\,\times\,$10$^{16}$
cm$^{-2}$ and the spacing is 0.2$\,\times\,$10$^{16}$ cm$^{-2}$.

    \ni  Fig.2~:~~Distribution  of  $^{12}$CO  kinetic temperature in L1616.
The elliptical outline of the I band  image of NGC 1788 is shown.  The  peak
contour level is 30$\,$K and the spacing is 2$\,$K.

\ni  Fig.3~:~~Distribution  of  equivalent  width  obtained  from $^{12}$CO
spectra.  The innermost contour level is 3.1 kms$^{-1}$ and the spacing  is
0.2 kms$^{-1}$.   The outer  open contours  are possibly  artefacts arising
from the uncertainty in the measured widths owing to poorer signal to noise
ratios.  The  portions of  the cloud  material lying  between the  open and
closed contours have linewidths  $\gsim\,2\,$kms$^{-1}$, but are not  shown
for the  sake of  clarity.  The  ellipse outlines  the I  band image of NGC
1788.

    \ni  Fig.4~:~~Distribution  of  integrated  $^{12}$CO  line intensity in
L1616.  The I band image of NGC  1788 is outlined by the ellipse.  The  peak
contour level is 33 Kkms$^{-1}$ and the spacing is 2 Kkms$^{-1}$.

    \ni  Fig.5~:~~The  {\it  thin}   and  {\it  thick}  contours   show  the
distribution of $^{12}$CO emission averaged over the velocity ranges 5.4  to
6.7 kms$^{-1}$ and 8.7 to  10.4 kms$^{-1}$, respectively.  The elliptical  I
band image of NGC 1788 is also marked.  The peak contour level is 5 K with a
spacing of 1 K.

    \ni Fig.6~:~~Bright Orion stars near the cloud L1616.  Its cometary tail
is indicated.  The details of the stars are given in Table 2.  The tail when
extended points  to stars  3 and  4.  Star  3 is  $\epsilon$ Orionis, a blue
supergiant.  Owing to its nearness  to the cloud and earlier  spectral type,
progenitor of $\epsilon$ Orionis is  more likely to be responsible  for both
the morphology and the efficient star formation of the cloud L1616.

    \ni Fig.7~:~~Shows the contour plot of the time taken by the cloud L1616
to  travel  from  its  initial  position  to  its  present  one, taken to be
50$\,$pc, as  a function  of the  initial distance,  $x_{0}$, and the escape
velocity  of  the  ionised  gas,  $v_{i}$,  when  accelerated  by  the  star
$\epsilon\,$Orionis through the rocket  effect.  Contours are labelled  with
the time taken in units of  million years.  The parameter space between  the
4.2$\,$Myr and  5.1$\,$Myr contours  is favourable  for the  star being  the
external trigger.

    \ni Fig.8~:~~Shows the time evolution of the cloud parameters as it gets
accelerated through  the rocket  effect.  In  each of  the graphs,  thin and
thick curves correspond to two cases with $v_{i},\,x_{0}\,{\rm  and}\,x_{1}$
of  12$\,$(15)  kms$^{-1}$,  5$\,$(4)  pc  and  50$\,$(70) pc, respectively.
Distance from the  star, ion density  just behind the  shock front, velocity
and mass expressed as a fraction of its present value, 180 M$_{\odot}$,  are
the respective parameters plotted in the four graphs.

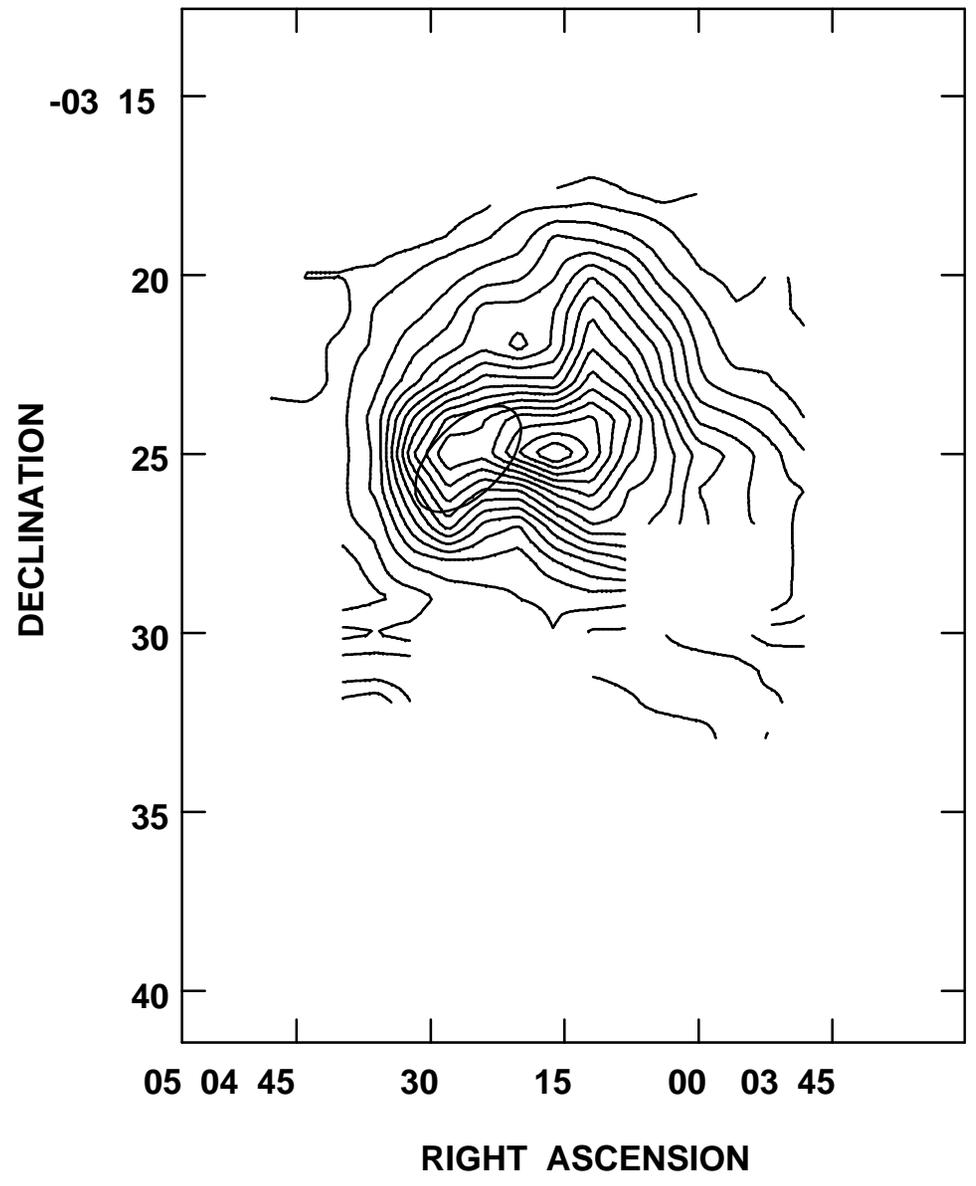
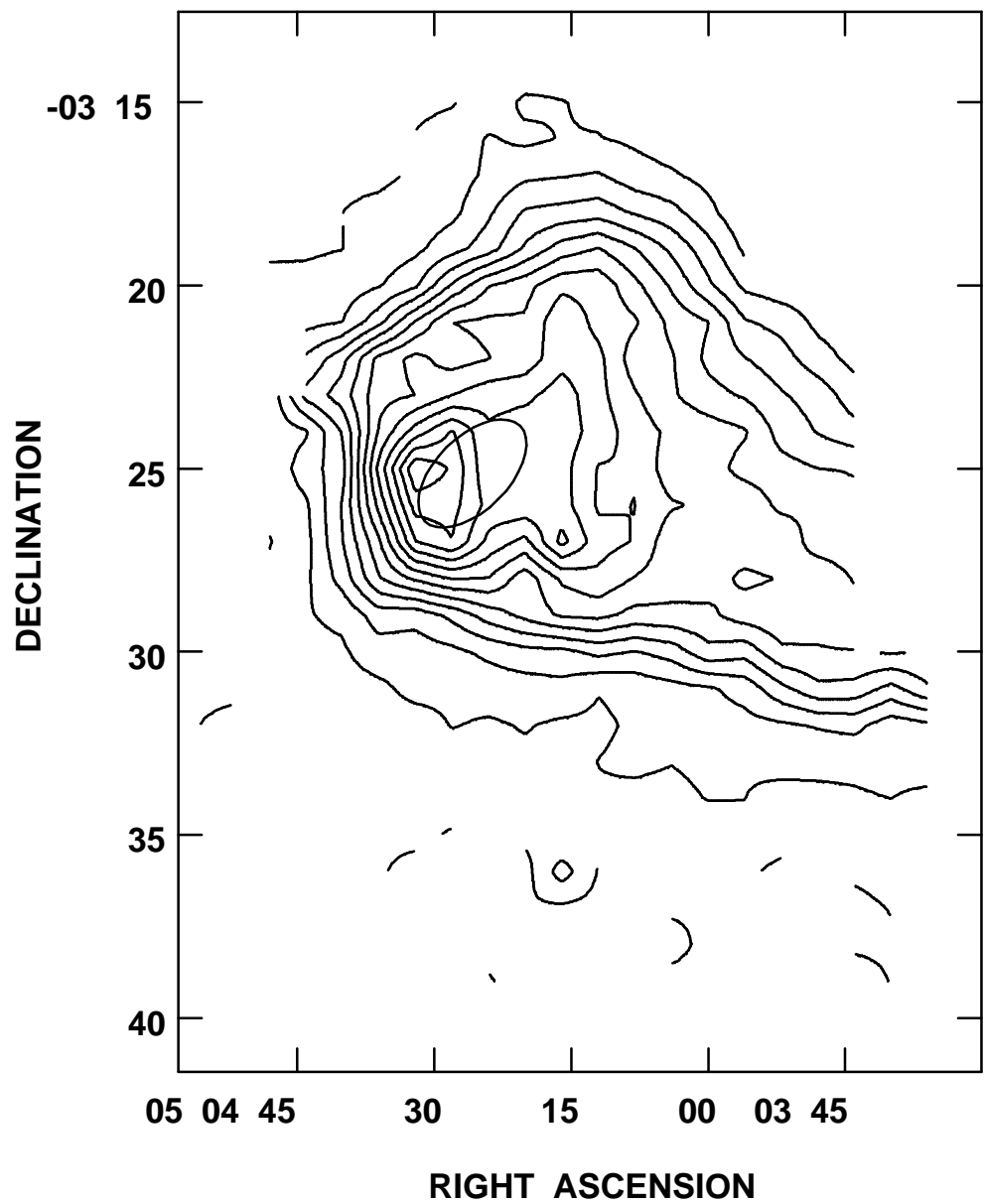
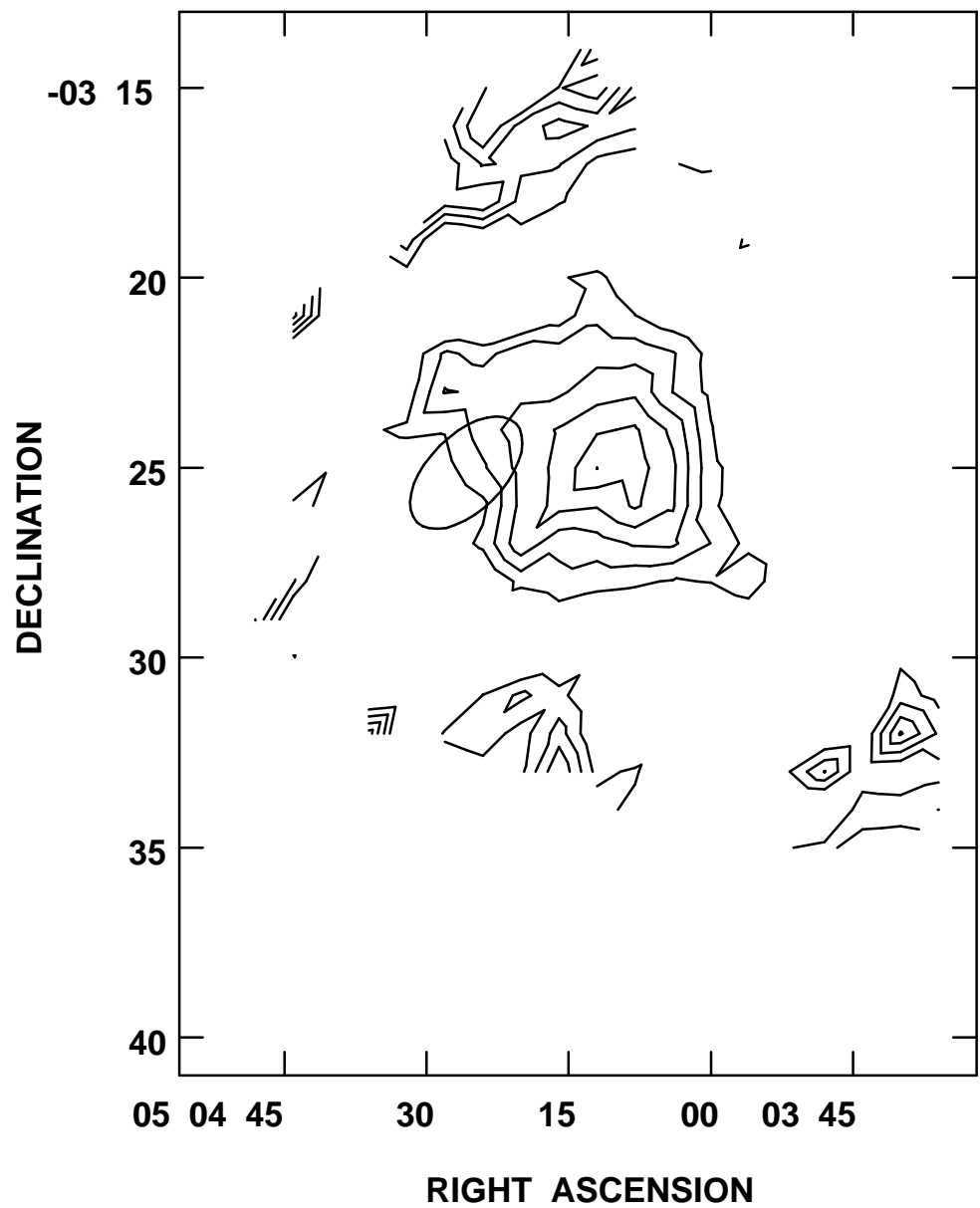
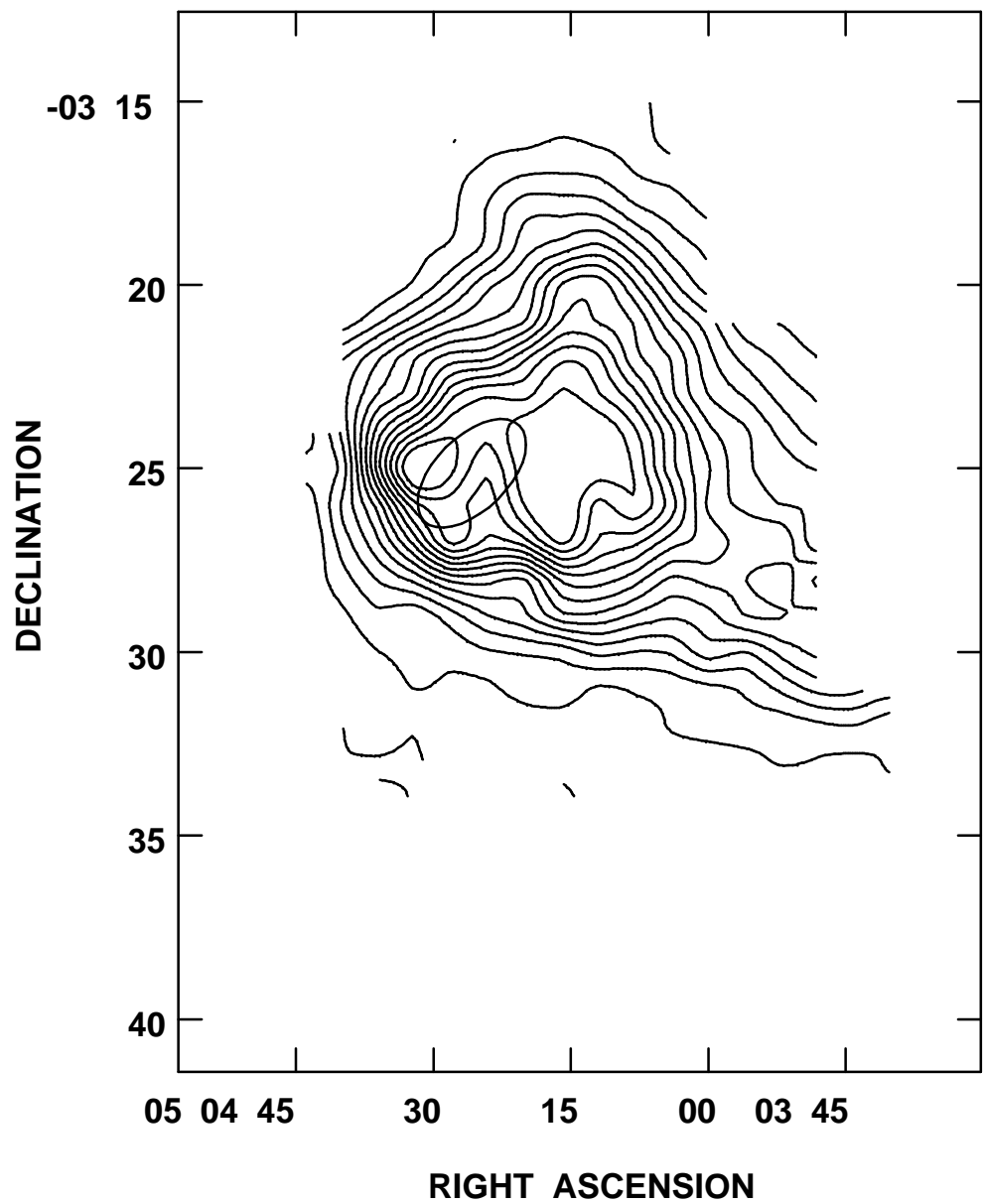
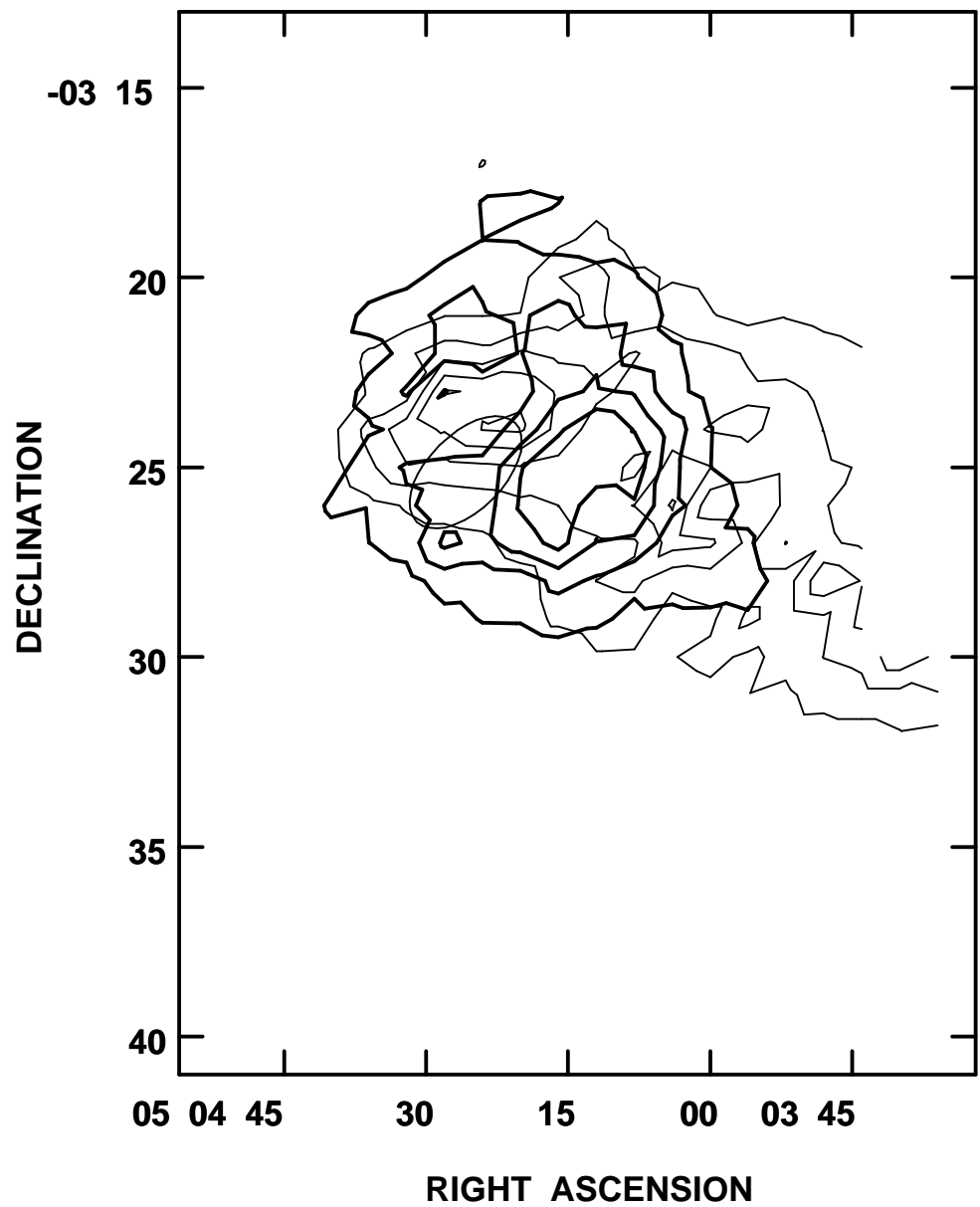
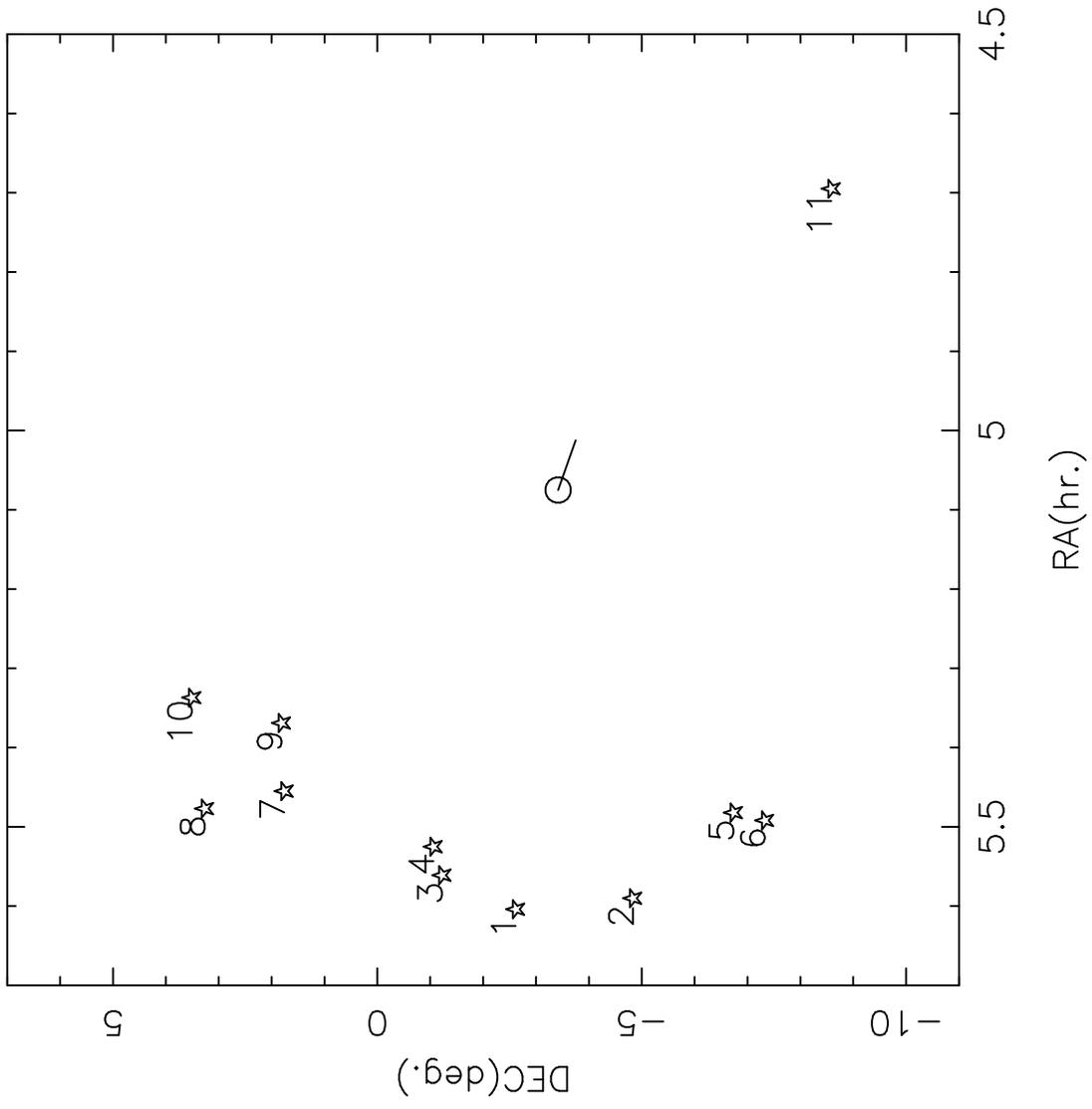
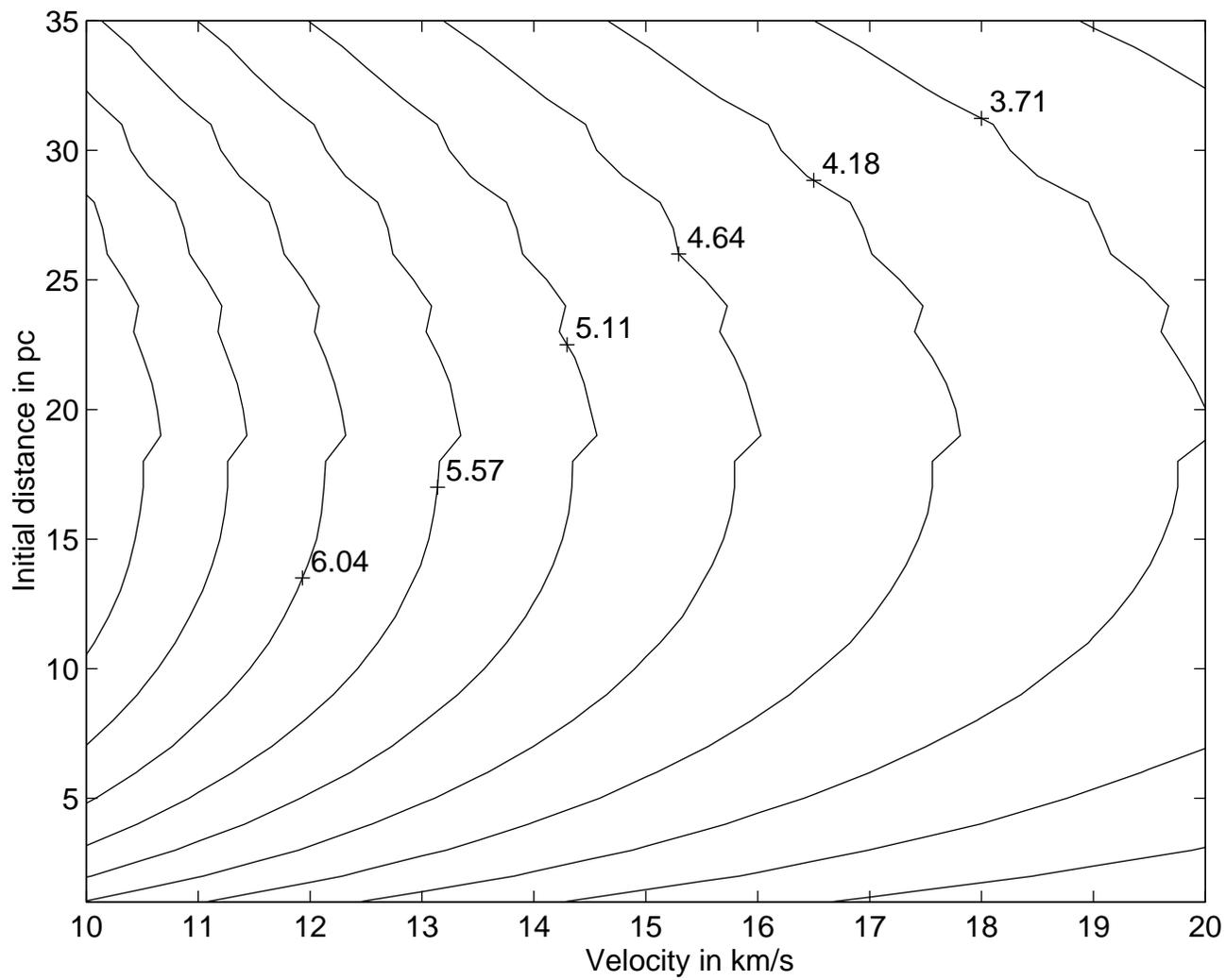
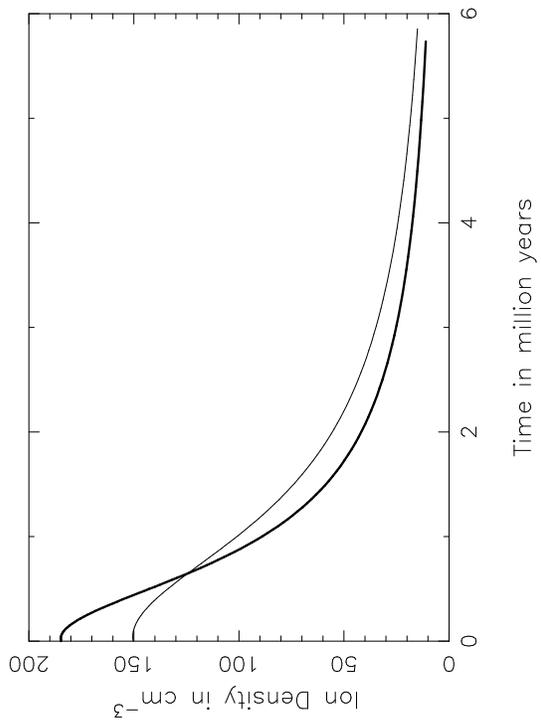
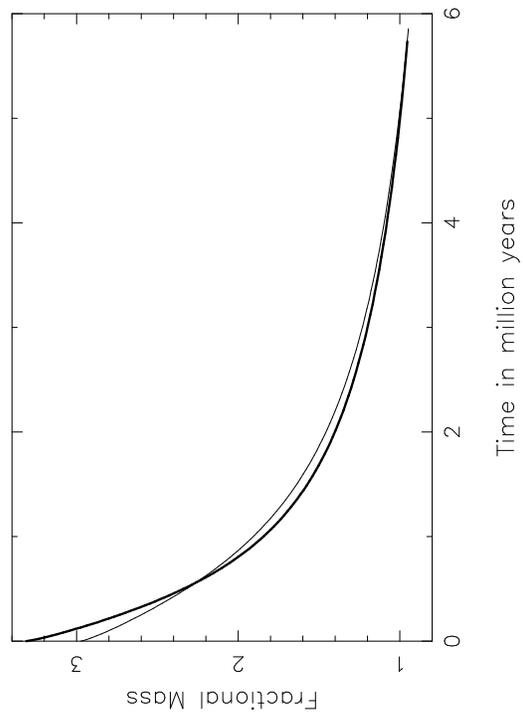
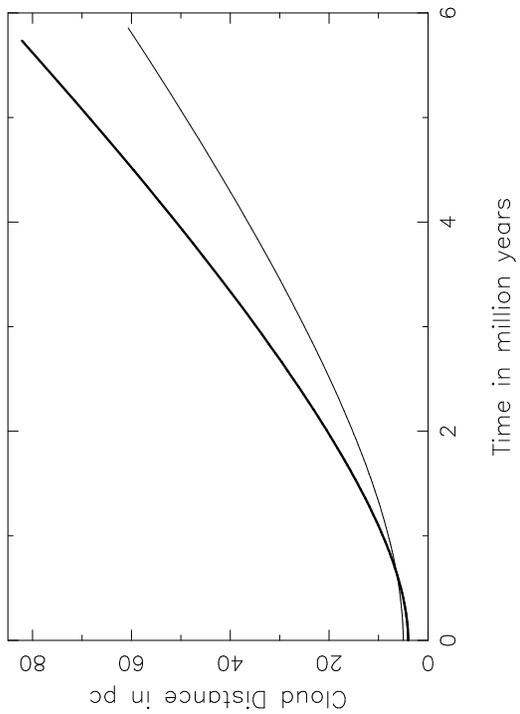
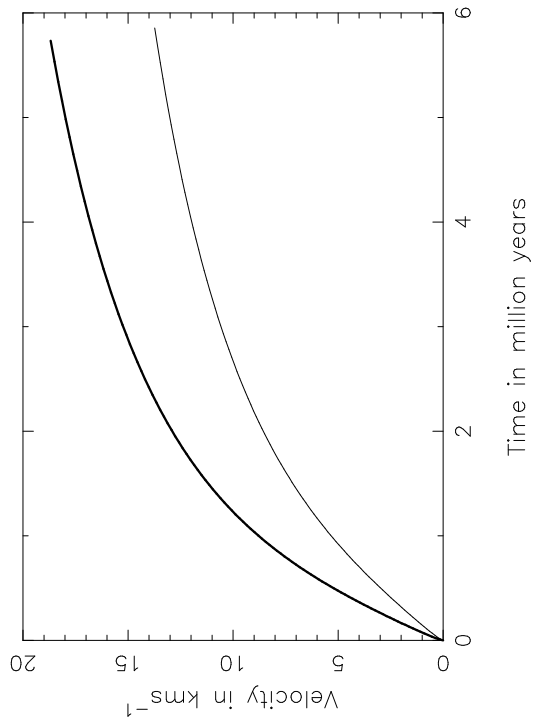
\end{document}